\documentstyle[12pt,epsf]{article}
\def\be{\begin{equation}}
\def\ee{\end{equation}}
\def\bea{\begin{eqnarray}}
\def\eea{\end{eqnarray}}

\def\br{\begin{thebibliography}{abc}}
\def\er{\end{thebibliography}}

\def\bar#1{\overline{#1}}

\def\real{\relax{\rm I\kern-.18em R}}
\def\com{\relax\,\hbox{$\inbar\kern-.3em{\rm C}$}}

\def\br{{\bf R}}

\def\inbar{\,\vrule height1.5ex width.4pt depth0pt}
\def\IC{\relax\,\hbox{$\inbar\kern-.3em{\rm C}$}}
\def\ID{\relax{\rm I\kern-.18em D}}
\def\IF{\relax{\rm I\kern-.18em F}}
\def\IH{\relax{\rm I\kern-.18em H}}
\def\II{\relax{\rm I\kern-.17em I}}
\def\I1{\relax{\rm 1\kern-.28em l}}
\def\IN{\relax{\rm I\kern-.18em N}}
\def\IP{\relax{\rm I\kern-.18em P}}
\def\IQ{\relax\,\hbox{$\inbar\kern-.3em{\rm Q}$}}
\def\IZ{\relax\,\hbox{$\inbar\kern-.3em{\rm Z}$}}
\def\R{\relax{\rm I\kern-.18em R}}
\font\cmss=cmss10 \font\cmsss=cmss10 at 7pt
\def\Z{\relax\ifmmode\mathchoice
{\hbox{\cmss Z\kern-.4em Z}}{\hbox{\cmss Z\kern-.4em Z}}
{\lower.9pt\hbox{\cmsss Z\kern-.4em Z}}
{\lower1.2pt\hbox{\cmsss Z\kern-.4em Z}}\else{\cmss Z\kern-.4em
Z}\fi}

\def\bar#1{\overline{#1}}

\def\Hat#1{\rlap{\kern.10em$\widehat{\phantom G}$}#1}
\def\HAt#1{\rlap{\kern.05em$\widehat{\phantom G}$}#1}

\def\czp#1{\rlap{\kern.1em$\widehat{\phantom{G\vrule height.8em}}$}#1{}}
\def\Czp#1{\rlap{\kern.05em$\widehat{\phantom{G\vrule height.8em}}$}#1{}}

\newcommand{\sect}[1]{\setcounter{equation}{0}\section{#1}}

\def\sxn#1{\bigskip\medskip \sect{#1} \smallskip
                                                 }

\begin{document}

\thispagestyle{empty}
\setcounter{page}{0}


\begin{flushright}
SU-4240-664\\
August 1997
\end{flushright}

\vglue 0.6cm

\centerline {{\Large{\bf Second Stage String Fragmentation 
Model}}}
 
\vglue 0.6cm
                        
\centerline {{\large  Thamar Ali 
Aithan\footnote{Current address: Physics Department, KFUPM, Dhahran-31261,
Saudi
Arabia}
 and Carl
Rosenzweig}}

 \vglue 0.5cm
{\centerline{\it {Department of Physics, Syracuse University,}}
\centerline{\it {Syracuse, NY 13244-1130, USA}}

\vglue 4cm
\centerline {\bf Abstract}
\vglue 0.6cm

A string model, advocated by Bowler, provides a physical and intuitive
picture of heavy quark fragmentation.  When supplemented by an ad hoc
factor of $(1-z)$, to suppress fragmentation near $z=1$, it supplies an
excellent fit to the data.  We extend Bowler's model by accounting for the
further  decay of the massive mesonic states produced by the initial string
breaking.  We find that each subsequent string break and cascade decay
beyond the first, introduces a factor of $(1-z)$.  Furthermore we find
that
including a finite mass for the quarks, which pop out of the vacuum and  
split
the string, forces the first string breaking to produce massive
states requiring further decay.  This sequence terminates at the second
stage of fragmentation where only relatively ``light'' heavy meson systems
are formed.  Thus we naturally account for the phenomenologically  required
factor of $(1-z)$.  We also predict that the  ratio of (primary) 
fragments-vector/(vector  plus scalar)  should be .61.  Our second stage
string
fragmentation model provides an appealing picture of heavy quark
fragmentation.

 \newpage
\vglue 0.6cm
\sxn{ {\bf   Introduction}}

The idea that QCD field configurations resembling a string, or flux tube, 
are
important in regimes  where confinement dominates is 
an  old and appealing one.  It receives its strongest support from its
birthplace -- the description of the spectrum of linearly rising Regge
trajectories \cite{su1}.  Its scope was greatly expanded by the proposal
of a
simple model of string breaking \cite{su2}. The string model
 is the basis
for the widely
employed  Lund model of fragmentation \cite{su3}. The application to
fragmentation processes has however, a glaring flaw.  Data on heavy meson
fragmentation clearly indicate that production of heavy mesons near $z=1$
is
highly suppressed, probably like $(1-z)$. The Lund model allows for 
$(1-z)^a$ behavior, but $a$ is undetermined.  Bowler \cite{su4} using a
string model originally proposed by Artru and Mennessier \cite{su5}
derived a string fragmentation model for heavy quarks with no $(1-z)^a$
suppression factor.  Bowler's distribution {\em peaks} at $z=1$.  If a
factor of $(1-z)^\beta$ is arbitrarily appended to Bowler's fragmentation
function an excellent fit to the data is obtained with 
$\beta=.95 \pm 0.11$ \cite{su6}.  

In this paper we propose an origin for the 
$(1-z)$ factor in fragmentation. It arises from allowing the string state
to continue fragmenting even after the initial  heavy mesonic system has
formed. The
continuing cascade decays carry away momentum and deplete the population
of high $z$, heavy mesons.   A crossing of a light and heavy  quark
worldline naturally defines a series of stages in this cascade decay.
Bowler ended his fragmentation at the first crossing.  We find that the
$(1-z)$ factor comes from the fragmentation  {\em after} this
crossing.  We refer to such a process as second stage fragmentation and in
Section III we present a derivation of the  corresponding fragmentation
function.  This second stage fragmentation function is phenomenological
equivalent to the modified Bowler function with the benefit that 
$\beta=1$
is naturally selected by the physics.  The second stage process is
distinguished from the modified Bowler function in having a sharper drop
off with hadron mass.  This leads to a prediction for the fraction of
vector to scalar plus vector production of .61 which differs from a value
of .67
for Bowler's modified function.

The second stage fragmentation function is thus seen to be an entirely
satisfactory description of heavy quark fragmentation.  What is special
about the second stage that the bulk of fragmentation should occur here?
In an attempt to answer this question we consider effects of finite mass, 
light quarks and explore a possible weakness of the tunneling model of
string
breaking.  A consequence of tunneling in the string rest frame (lab frame 
for the first stage) with finite mass, light quarks  is that the mesonic
state produced is very massive. Its mass is significantly greater than the
mass of the
lightest ``stable'' charm mesons. These mesonic states will therefore
further fragment i.e. continue their fragmentation into the second stage.
The probability of producing a $D_0$ or $D^*$ at the first stage is small
because of the kinematics of the light quark mass. When
we apply the same
reasoning at the second stage, the first stage dynamics and the second
stage  kinematics is such that only ``light'' heavy quark mesons are 
produced.  These ``light'' states do not need to undergo further
fragmentation.

The second stage  seems to be special.  First stage fragmentation is
unlikely to be the final stage. It  leads to a massive fireball
which further fragments, whereas the second stage is not kinematically
forced to continue fragmenting.  We do not claim that this is a
complete answer, but rather that it is indicative of what might be special
about second stage fragmentation.  Our arguments on this ``specialness''
are presented in section IV.

\sxn{{\bf First Stage Fragmentation}}    

In this section we review the motion and fragmentation of a heavy quark,
-anti-quark pair $(Q,\bar{Q})$ produced  in  $e^+e^-$ annihilation, and
bound
together
by a string.  Although the kinematics are somewhat simpler in the heavy
quark rest frame, for our purposes it will prove useful to work in the
laboratory frame.  The $Q\bar{Q}$ are produced with initial momentum 
$P_0$, equal to the electron laboratory frame momentum.  The equation of
motion of $Q$ (chosen  to be moving to the right) is
\be
\frac {dP}{dt} = \frac {d}{dt} \left ( \frac {\mu v}{\sqrt {1-v^{2}}}
\right )
=-\alpha \label{2.1}
\ee
where $\alpha$ is the string tension and $\mu$ is the heavy quark mass.
The solutions to the equations of
motion are
\be
P_Q(t)=P_0-\alpha t \label{2.2}
\ee
\be
\alpha x_Q(t) = \sqrt{\mu^{2} + P_{0}^{2}} -
\sqrt{\mu^{2}+(P_{0}-\alpha t)^{2}}. \label{2.3}
\ee
It will prove convenient to use light cone variables
\be
x^+\equiv \frac {(x+t)}{\sqrt{2}} ~~~~~~x^-\equiv \frac{(t-x)}{\sqrt{2}}
\label{2.4}
\ee
in terms of which the equation of motion (\ref{2.3}) is 
\be
x^+= \frac {x^{-}(P_{0}+\sqrt{P^{2}_{0}+\mu^{2})} }
{\sqrt{2 }\alpha x^{-}  + \sqrt {P^{2}_{0} + \mu^{2}}-P_{0} }
\label{2.5}
\ee
or  equivalently
\be
x^- = \frac {(\sqrt{\mu^{2}+P^{2}_{0}} - P_{0})x^{+}}
{\sqrt{\mu^{2}+P^{2}_{0}}+P_{0}- \sqrt{2} \alpha x^{+} }. \label{2.6}
\ee

The  string joining $Q$ with $\bar{Q}$ will eventually break at some point
$(x_1,t_1)$ (or equivalently $(x_{1}^{+},x_{1}^{-})$) producing  a
$q_1\bar{q}_{1}$ pair, which for the moment, we take to be massless.  The
$\bar{q}_{1}$  and $Q$ trajectories will intersect at point
$(x_{m_{1}}^{+}, x_{m_{1}}^{-})$ signaling the formation of a heavy 
quark mesonic system, or   fireball,  of mass $m_1$. This terminates what
we
shall refer to as
first stage fragmentation (see Fig. 1). The $\bar{q}_{1}$  acquires 
both
momentum and energy from being accelerated by the string connecting it to
$Q$ so that when it intersects $Q$
\be
P_{\bar{q}_{1}}= \alpha(t_{m_{1}} -
t_1)~~~~~
E_{\bar{q}_1}=\alpha(x_{m_{1}}-x)
\label{2.7}
\ee
whereas
\be
P_{Q}(t_{m_{1}}) = P_0 - \alpha t_{m_{1}} ~~~E_Q (t_{m_{1}})
 = \sqrt{\mu^{2}+P_{0}^{2}} - \alpha x_{m_{1}} . \label{2.8}
\ee
Thus the energy and momentum of meson $m_1$ are
\be
P_{m_{1}} = P_0-\alpha t_1~~~~~E_{m_{1}} = \sqrt{\mu^{2}+P_{1}^{2}}-\alpha
x_1 \label{2.9}
\ee
or
\be
(E+P)_{m_{1}} = P_0+ \sqrt {\mu^2+P_{0}^{2}} - \sqrt {2} \alpha  x^+
\label{2.10}
\ee
and
\be
(E-P)_{m_{1}}= \sqrt{\mu^2+P_{0}^{2}}-P_0 + \sqrt{2}\alpha x^{-} .
\label{2.11}
\ee
The kinematic invariants we are interested in are the heavy meson mass 
$m_1$ and $z_1$ the  fraction of the total momentum
carried by it. 
\be
z_1\equiv \frac {(E+P)_{m}}
{\max (E+P)_{Q}}=
\frac {(E+P)_{m}}
{P_{0}+\sqrt{\mu^{2}+P_{0}^{2}}} \label{2.12}
\ee
These are determined by the light cone variables $(x_{1}^{+},
x_{1}^{-})$
as 
\be
z_1 = 1 - \frac {\sqrt{2} \alpha x_{1}^{+}}
{P_{0}+\sqrt{\mu^{2}+P_{0}^{2}} } \label{2.13}
\ee
\be
m_{1}^{2}=z_1(\mu^2+(P_0+\sqrt{\mu^{2}+P_{0}^{2}})\sqrt{2}\alpha
x_{1}^{-})). \label{2.14}
\ee

The probability that $m_1$ will form at $x_{m_{1}} t_{m_{1}}$
is the probability that no other break occurred  in the absolute part of 
$(x,t)_{m_{1}}$.  Any such break would pre-empt $m_1$ and lead to a
different heavy mesonic system.
This probability  of formation of $m_{1}$ is
\be
dP=\rho d A e^{-\rho A}
\label{2.15}
\ee
$\rho$ is the constant probability, per unit 4 space-time volume, that
the string will break in $dxdt$.  $A$ is the area 
 in the absolute past of $m_1$. A mildly tedious but straightforward
calculation leads to 
$$
dA = \frac {dm_{1}^{2}}
{z_{1}}
~\frac {dz_{1}}
{2\alpha ^{2}}
$$
\be
dP=\frac{\rho}
{2\alpha^{2}} 
d m_{1}^{2}
\frac {dz_{1}} {z_{1}} 
\exp \left \{ 
\frac{-\rho\mu^{2}}
{2\alpha^{2}}
\left [ \left ( \frac {m_{1}}{\mu}
\right )^2
\frac {1}{z_{1}} - 1 -\ln \left [ \left( \frac {m_{1}}{\mu}\right )^2
\frac{1}{z_{1}}
\right ] \right ] 
\right \}
\label{2.16}
\ee
implying the first stage fragmentation function
\be
f(m_{1}^{2}, z_1) = B\frac{1}{z_{1}}\exp 
\left \{-B\mu^2 \left [ \left ( \frac {m_{1}}{\mu}
\right )^2
\frac {1}{z_{1}} - 1 -\ln \left [ \left ( \frac {m_{1}}{\mu}\right )^2
\frac{1}{z_{1}}
\right ] \right ] 
\right \}
\label{2.17}
\ee 
where 
\be
B\equiv \frac{\rho}{2\alpha^{2}}.
\label{2.18}
\ee
This is the well known result of Bowler.

For $z_1$ close to one $((1-z_1))\ll 1$,
\be
\int f(m_{1}^{2} z_1)dm_{1}^{2}\approx
\sqrt{\frac {\pi B\mu^{2}}{2}}
- \frac {B\mu^2}{z_{1}} (1-z_1).
\label{2.19}
\ee
while for larger $z_1$ values
\be
\int f(m_{1}^{2}, z_{1})dm^{2}_{1} \approx \frac {z_{1}}{1-z_{1}}e^{-\frac
{B\mu^{2}}{2}\left (\frac {1-z_{1}}{z_{1}}\right )^{2}}.
\label{2.20}
\ee
\sxn{{\bf  Second Stage Fragmentation}}
It will prove kinematically convenient to perform this calculation in the
initial rest frame of the heavy quark.  We refer to Figure 2 for the
definition of the important spacetime points involved in the
fragmentation.  The heavy quark $Q_0$ is formed at the origin.
A $q_1\bar{q}_1$ pair forms at $(x_1 t_1)$ and the $\bar{q}_1$ crosses the
$Q_0$ trajectory at $m_1$. Between $t_1$ and $t_{m_{1}}$ the $q_1 Q$
system is an  extended mesonic system or fireball.  It is not necessarily
a
stable or even semi-stable meson.  Although the actual crossing point
$m_1$ is of little physical significance it is a convenient distinguishing
point.  Fragmentations occurring before $t_{m_{1}}$ we refer to as first
stage fragmentation.  The events occurring between $t_{m_{1}}$
and $t_{m_{3}}$
are referred to as second stage fragmentation.  Higher order
stages are defined similarly.  During the second stage process another
$q\bar{q}$ tunnel into existence at $(x_2, t_2)$.  This leads
to the observed heavy meson fireball forming at $m_3$, while another light
quark state forms at $m_2$.

The energy-momentum of $m_3$ is the sum of the energy-momentum of $Q_0$ at
space-time point $(m_3)$, and the energy-momentum of $\bar{q}_2$.  The
energy-momentum of $Q_0$ at space-time point $(m_3)$, evaluated in the
initial rest frame of $Q_0$, is
\be
E_{Q_{0}}(m_3) = \mu + \alpha  (x_{m_{3}}-2x_{m_{1}})
\label{3.1}
\ee
\be
P_{Q_{0}}(m_3) = \alpha(t_{m_{3}}- 2t_{m_{1}})
\label{3.2}
\ee
 We have used as a starting point the energy-momentum of $Q_{0}$ at
point $m_1$.
The energy-momentum of $\bar{q}_2$ is 
\be
E_{\bar{q}_{2}}(m_3) = \alpha(x_2-x_{m_{3}})
\label{3.3}
\ee
\be
P_{\bar{q}_{2}}(m_3) = \alpha(t_2-t_{m_{3}})
\label{3.4}
\ee
leading to the energy-momentum of meson $m_3$
\be
E_{m_{3}} = \mu + \alpha(x_2-2x_{m_{1}})
\label{3.5}
\ee
\be
P_{m_{3}} = \alpha(t_2-2t_{m_{1}})
\label{3.6}
\ee
The relevant kinematic invariants, are $z_3$ the fractional momentum
carried
by the heavy meson $m_3$ and  its mass squared $m_{3}^{2}$.
These can be expressed in the coordinates of Fig. 2 as
\be
z_3 \equiv \frac {E_{m_{3}} + P_{m_{3}}}
{\mu}
= 1+ \frac{\sqrt {2}\alpha}
{\mu}
(x_{2}^{-}-2x_{m_{1}}^{-})=
z_1-\frac {\sqrt{2}\alpha}{\mu}(x_{m_{2}}^{-}-x_{2}^{-})
\label{3.7}
\ee
\be
m_{3}^{2}\equiv (E_{m_{3}}+P_{m_{3}})(E_{m_{3}}-P_{m_{3}}) = 
\mu^2 z_{3}(1-\frac{\sqrt{2} \alpha}{\mu}(x^{+}_{2}-2x_{1}^{+}))
=\frac{m_{1}^{2}z_{3}}{z_{1}}- 
 \frac{m_{2}^{2}z_{3}}{z_{2}}.  
\label{3.8}
\ee
The last expression for $m^{2}_{3}$ is easily derived using the
conservation of energy-momentum $P_\mu, P_{m_{3}}^{\mu}= P_{m_{1}}^{\mu}-
P_{m_{2}}^{\mu}.$  

The probability $dP_2$, of forming a meson at $m_3$
in the second stage, is

 $dP_2=$ (Probability of forming
 $m_1$) $\cdot$ (probability of no
break
in
absolute past of $m_3$)
\be
=(\rho d A_1.e^{-\rho A_{1}}).(\rho d A_{2}.e^{-\rho A_{2}})\equiv
dz_1 dm_{1}^{2}dz_{3} dm_{3}^{2}f(z_1,m_{1}^{2}, z_3,m_3)
\label{3.9}
\ee
so
\be
f(z_1,m_{1}^{2}, z_3, m_3)=
\left ( \frac {\rho} {2\alpha^{2}}\right )^{2} 
\frac {e^{-\rho(A_{1}+A_{2})}} {z_{1}z_{3}}
\label{3.10}
\ee
The calculation of the second stage fragmentation function is now reduced
to the geometrical calculation of the area $A_2$.
In order to calculate $A_2$ we need the equation of motion for $Q_1$ after
it passes the crossover point $m_1$.
\be
x^{+}(x^{-})= \frac {\mu^2}{2\alpha^{2}[2x_{1}^{-}+\mu-x^{-}]}+2
x^{+}_{m_{1}}-
\frac {\mu}
{\sqrt{2} \alpha}
\label{3.11}
\ee
valid in the range
\be
x_{1}^{-} \leq x^{-} \leq x_{m_{3}}^{-}
\label{3.12}
\ee
Recall we are now in the heavy quark initial rest frame, where
\be
x_{m_{1}}^{+}=
\frac {\mu x_{1}^{-}}
{(\mu + \sqrt{2}\alpha x_{1}^{-})}=
\frac {\mu}{\sqrt{2} \alpha}
\left \{ 1- \frac {\mu^{2}z_{1}}
{m_{1}^{2}}  \right\}.
\label{3.13}
\ee
We also need
\be
x^{+}_{2} = \frac{\mu}{\sqrt{2} \alpha}
\left\{z_3 + 1 - 2 \frac{\mu^{2}z_{1}}{m^{2}_{1}}
\right\}
\label{3.14}
\ee
and
\be
x_{3}^{-} = \frac {\mu} {\sqrt{2}\alpha}
\left ( \frac {2m_{1}^{2}}{\mu^{2}z_{1}} -1 - \frac{1}{z_{3}}   \right ).
\label{3.15}
\ee
Putting this together with the geometry of the shaded area in Fig. 2 we find
\be
A_{2}= \frac{\mu^{2}}{2 \alpha ^{2}}
\left [ \frac {m_{1}^{2}}{\mu^{2}z_{1}} z_{3}-
1-\ln \left ( \frac {m_{1}^{2}} 
{\mu^{2}z_{1}}z_{3} \right )   \right ].
\label{3.16}
\ee
It is notationally convenient to express the fragmentation function as a
function of the variables 

$$X_{1} \equiv \frac{\rho}{2\alpha^{2}}~ \frac{m_{1}^{2}}{z_{1}}=
\frac{Bm_{1}^{2}}{z_1}$$
 and
$$X_0=B\mu^{2}$$
\be
f(z_3,m_{3}^{2},z_{1},X_{1}) =
\frac{B^{2}}{z_{3}z_{1}} 
\left( \frac{X_{1}^{2}z_{3}} {X_{0}^{2}}\right )^{X_{0}}
e^{2 X_{0}-X_{1}(1+z_{3})}.
\label{3.17}  
\ee
The experimentally relevant function involves only the mass $(m_3)$ and
fractional momentum $(z_3)$ carried by the ``meson'' $m_3$.  It is
obtained by integrating over $z_1$ and $m_{1}^{2}$
\be
f(z_3,m^{2}_{3}) = \int \int dz_1 dm^{2}_{1} f(z_3,m^{2}_{3}, z_{1},
X_{1}). 
\label{3.18}
\ee
 We now need the limits of integration.  For fixed $z_3$  and $ m_{3}^{2}$
$$
z_3 < z_1 < 1
$$
$$
\frac {m_{3}^{2} z_{1}}
{z_3} < m_{1}^{2} < 4 P_{0}^{2} \approx \infty.
$$
Changing integration variables from $m^{2}_{1}$ 
to  $X_1$, and defining
$X_3\equiv \frac{Bm_{3}^{2}}{z_3}$
we find for the second stage fragmentation function 
\be
f(z_3, m_{3}^{2}) =
\frac{1}{z_3}\int_{z_{3}}^{1} dz_{1} 
\int_{X_{3}}^{\infty} dX_1 \left ( \frac{X_{1}^{2}}{X_{0}^{2}}
z_3
\right ) ^{X_{0}} 
e^{2X_{0}-X_{1}(1+z_{3})}
\label{3.19}
\ee
\be
= \frac{1-z_3}{z_{3}} \int_{X_{3}}^{\infty} 
 \left ( \frac{X_{1}^{2}z_{3}}{X_{0}^{2}}
\right )^{X_{0}} 
e^{2 X_{0} - X_{1}(1+z_{3})}
dX_1.
\label{3.20}
\ee
This is our key result.  A factor of $(1-z_3)$ naturally arises from the
string model. Physically it comes from the fact that the light meson,
$m_2$, produced in second stage fragmentation (see Fig. 2) carries off
part of the
initial momentum, leaving $m_3$ with a smaller fraction of $z_1$. The
integrand in 3.20 carries with it the sharp falloff in $\frac {1}{z_{1}}$
and $m_3$ that characterized first stage fragmentation.

It is illuminating to compare this second stage fragmentation function
to the first stage fragmentation by considering the limit $X_{3}>>X_0$ 
 (i.e. $\frac {m^{2}_{3}}{z_{3}}>>\mu^{2})$. 
 Equation (\ref{3.20})  can then be approximated as 
\be
\left( \frac {(1-z_{3})}{z_3}\left ( \frac {X_3}{X_0} \right)^{X_0}
e^{X_{0}-X_{3}}
\right ) \left [ \left (\frac {m^{2}_{3}}{\mu^{2}}  \right )^{X_{0}}
e^{B(\mu^{2}-m_{3}^{2})}
\left (1+ \frac{\mu^{2}}{m_{3}^{2}}
\left ( \frac {2 z_{3}}{1+z_{3}}   \right )
+...   \right )
\right ]
\label{3.21}
\ee

The first term in this expression is identical to the Bowler expression
with the {\em crucial} addition of a factor  $1-z_3$.  At fixed value of
$m^{2}_{3}$
the factor in square brackets is a weakly varying function of $z_3$ that 
changes
the normalization of the Bowler distribution. At fixed $z_3$ the square
brackets provides an additional fall off with $m_{3}^{2}$ when compared to
the Bowler function.

In fitting  the experimental $D^*$ and $D^0$ fragmentation functions a
modified form of the Bowler fragmentation function has been employed
\cite{su6}. A
factor  of $(1-z)^\beta$ was arbitrarily appended to the function of
(\ref{2.17}), and $\rho$ was treated as a parameter.  The so called
modified Bowler parametrization is
\be
\frac{(1-z)}{z}^{\beta} e^{B(\mu^{2}-m^{2})} \left (\frac{m^{2}}{\mu^{2}}  
\right )^{B\mu^{2}}
\label{3.22}
\ee
where $\beta$ was determined by fitting the data as $\beta=.95 \pm.11$

We propose the second stage string fragmentation function (\ref{3.20})
 as
a
substitute
for the modified Bowler form.  The $(1-z)$ factor, i.e. $(\beta=1)$ is a
direct consequence of the second stage fragmentation. The similarity
between Eqs. (\ref{2.17}) and (\ref{3.21}) generates an equally good fit.
In
Fig. 3 we compare the second stage function (\ref{3.20}) (with $B=.65$,
$\mu= 1.5$)  to the modified
Bowler function used in CLEO collaboration fits $(\beta=.95, B=.63)$. They
are indistinguishable.  The second stage fragmentation function  thus
provides a two parameter (overall
normalization factor, and $B$) fit to the full $z$ range of fragmentation
for both $D^0$ and $D^*$.

Given that $D^0$ and $D^*$ distributions are in accord with the 2nd stage
predictions we naturally ask whether their relative production rates are
given by our model. String models predict a suppression of large mass
states with respect to lighter states.  This is evidenced in the
exponential drop off with mass in Eqs. (\ref{2.17}) and (\ref{3.20}). (Even
though the mass difference between $D^*$ and $D^0$ is usually attributed
to
single gluon exchange, and not to any string dynamics our prediction of
exponential suppression in $m^2$ is robust.  This is because, whatever the 
origin of the extra mass,  in our model this  mass must be paid for
by a longer string
segment.  The energy needed to form more massive states has to come from 
the string energy.  How that piece of string re-arranges itself to form
the  final meson, is not relevant to the accounting of how much string
needs to be allotted to the state.  Therefore the mass dependency will be
valid since it depends only on how much string (energy) is transferred to
the produced meson.)

The experiments cannot distinguish between primordially produced $D^*$,
and those that are the end products of decay chains.  What can be
unraveled is the ratio.
\be
P_v \equiv \frac {N_v}{N_{v}+N_{s}}
\label{3.23}
\ee
where $N_{v(s)}$ refers to the number of primordial vectors (scalars)
produced.  Naively this ratio should be 3/4 since $N_v=3N_s$ from spin
state counting.  In string models the more massive vectors are produced
less
frequently than the lighter mass $D^0$, so $N_v < 3N_s$. Integrating the
second
stage fragmentation function (\ref{3.20}) over $z$ for both $D^0$ and
$D^*$ we
find
\be
P_v = .61
\label{3.24}
\ee

A similar prediction can be made for $D^{*}_{s}$
  to $D^{0}_{s}$ production, and we find an identical result
\be
P_v=.61
\label{3.25}
\ee

	The second stage fragmentation has a sharper drop-off in mass than
the modified Bowler form (\ref{3.22}).  Therefore the results of
(\ref{3.24}) and
(\ref{3.25}) are
smaller than we would get from (\ref{3.22}).  The modified Bowler
prediction is 
$$
P_v = .67
$$

The ratio $P_v$ thus offers an experimentally measurable distinction
between
the second stage fragmentation and the modified Bowler fragmentation
function.

In this section we have calculated the heavy quark fragmentation function
at the end of the second stage of fragmentation.  The heavy meson which
appears is the result of two events of string breaking.  The fragmentation
function  automatically incorporates the factor of $(1-z)$
which is required by experiment.  It is phenomenologically indistinguishable
from the modified Bowler function.  The experimental data thus strongly
support the string fragmentation model whereby heavy quark mesons are the
end result of a two stage, string breaking cascade decay.

We can continue the cascade process by considering a third stage of
fragmentation.  An additional factor of $(1-z)$ arises leading to a
$(1-z)^2$ behavior near $z=1$.  Even more rapid $m^2$ damping is also
present.  It appears the trend will continue for ever higher cascades.
Thus while the data could certainly accommodate a small admixture of third
or higher stages, the preponderance of fragmentation seems to come from
second stage string breaking.
\sxn{{\bf Finite, Light Quark Mass, Tunneling and Something Special about
the 2nd Stage}}
There are no massless quarks in nature.  The $u,d$ quarks have a
constituent quark mass of $\sim$ 300 MeV. We also know that hadron
fragmentation is not purely 1 space-dimensional.  Strings have finite
thickness and particles are produced with momentum transverse to the jet
direction.  It should be possible to incorporate some of these transverse
momentum effects into an effective mass.
\be
m_{eff}=\sqrt{m^{2}+<P^{2}_{\bot}>}
\equiv m_{\ell}
\label{4.1}
\ee

It therefore behooves us to re-examine the simple model of section II and
III for any significant effects that might arise from the  finite mass
$m_{\ell}$ of $q$.

One immediate consequence of finite mass is that a finite segment $\ell$,
of string, with $\ell=\frac{2m}{\alpha}$ must disappear in order for the
quarks to materialize when the string breaks.  There will be a gap between
the $q$ and $\bar{q}$ which are produced. We work in the laboratory rest
frame. (See Fig.4.)

We refer to the breaking point $x_1,t_1$, as the point midway between the
$q$ and $\bar{q}$ which have tunneled out of the vacuum, see Fig. 4.
Except if breaking occurs very near the turning point of $Q$, the
$\bar{q}$ will have ample time
for its trajectory to asymptote to the trajectory of a massless
$\bar{q}$
produced at $x_1$. This is indicated by the dotted line in Fig. 4.  
Therefore the intersection point $m_1$ is the same as
in section II for string breaking at $x_1$  and
\be
P_{m_{1}}= \alpha(t_{m_{1}}-t)~~~E_{m_{1}} =
\alpha(x_{m_{1}}-x_{\bar{q}})+m_\ell=\alpha (x_{m_{1}}-x_{{1}})
\label{4.2}
\ee
since $m_l=\alpha(x_{\bar {q}}-x_1)$.  The kinematics are identical to the
massless
case for breaking at $x_1$ where $x_1$ and $x_{\bar{q}}$ are defined  as
in Fig.
4.  An important difference is that $x_1$ must be a finite distance away
from the $Q$ trajectory.  It cannot get closer than $\Delta x \equiv 
\frac {m_\ell}{\alpha}$ and must therefore produce a mesonic system with
mass
$M>\mu$.  

The simplest way to find the minimum mass is to consider the
case where  $\bar{q}$, with mass $m_\ell$, materializes, at rest,
immediately
next to the moving $Q$
\be
M^2=(E_Q+m_\ell+P_Q)(E_Q+m_\ell
-P_Q)= \mu^2+2m_\ell E_Q+m^{2}_{\ell}\sim \mu^2+2m_{\ell}E_Q.
\label{4.3}
\ee
$z$ for this state is 
\be
z=\frac{E_{Q}+P_{Q}+m_{\ell}}{P_{0}}
\label{4.4}
\ee
implying 
\be
M^2= \mu^{2}+2m_\ell\sqrt{P^{2}_{0}z^{2}+\frac{\mu^{2}}{2}}
\label{4.5}
\ee

 Except for very small $z$, $M$ will be large,  significantly
greater than $\mu^2$.  

Large mass states, in general, are unstable and
readily decay, emitting pions until the lowest mass, heavy quark meson is
produced.  Thus the fragmentation considered by Bowler and described in
section II is not the final result, but the first step in a cascade
process.  Only after several steps or stages will a relatively stable,
heavy quark meson be produced terminating the  cascade and producing the
experimentally observed fragment.

To make these observations concrete consider  charm quark fragmentation
at CESR and ARGUS with $P_0 \simeq \frac {10.55}{2}$ GeV/c, $\mu= 1.5$
GeV.

We chose $m_\ell \simeq
350 - 400$ MeV, which  is a reasonable value for the constituent quark
mass described in the discussion surrounding (\ref{4.1}).  We now ask;
for what values of $z$  will the fireball produced by the (first stage)
fragmentation of Fig. 4 continue to decay?  We take the minimal mass $M$  
of
this fireball to be $2.15$ GeV. This is above $(m_\pi + m_{D^{*}})$
providing phase space to decay into the ``stable'' minimal mass heavy
charmed mesons $D^0(1869)$ and $D^*(2010)$.  From Eq. (\ref{4.5})
\be
z \geq \frac{\sqrt{\left( \frac{M^{2}-\mu^{2}}{2m} \right)^{2}-
\frac{\mu^{2}}{2}}}{P_{0}}
\simeq .52-.61
\label{4.6}
\ee
depending on our choice of $m_\ell$.  This result is also sensitive to the
value of $\mu$, varying from $.44$  to $.52$ for $\mu = 1.6$ GeV. 

 For all $z$ {\em greater than} .5 to .6, {\em massive} fireballs will  be
produced, with mass $\geq$ 2.15 GeV.  Such massive states require 
further cascade decays before producing stable
states. Since the original, first stage, fragmentation function is
strongly peaked near $z=1$, (see Eq. (\ref{2.17}) and (\ref{2.19}))
between 95\%  and 75\% of all states produced by first stage string
breaking will proceed to  second or higher stages. Only the 5-25\%
produced with small $z$ $(z \leq .5~\rm{to}~ .6)$ will be observable as
$D$ or
$D^*$ products of first stage fragmentation. All high $z~D_0$ or $D^*$ 
must be the result of second, or higher, stage fragmentation and will be
described by e.g. (\ref{3.20}), or its higher interations.  The
finiteness of the light quark mass and the kinematics of string breaking
force us into a multi-cascade picture of fragmentation. As we showed in
Section 3  cascade decay
automatically suppresses fragmentation near $z=1$.

Our result for the minimum mass of the fireball is strongly frame
dependent since it arises from the relative momentum of $Q$ with respect
to the $\bar{q}$ which has tunneled from the vacuum.  The tunneling
phenomena itself is frame dependent and seems to us most reliable in  the
string rest frame.

  Since the
quark and  anti-quark which tunnel out of the vacuum have mass $m$, a
finite segment of string $\ell$, determined by $\ell=2\frac {m} {\alpha}$
will disappear.  This phenomena produces the quark mass dependence 
$e^{-m}$  characteristic of tunneling.  It also provides problems for the
tunneling model.  The quark and  its anti-quark partner will materialize
at different space-time points, hence the tunneling is non-local.  (By
contrast, massless quarks will always pop out simultaneously at the same
spot.  Thus string breaking by massless quarks is local.) An immediate
consequence of this is that the tunneling looks dramaticly different in
different Lorentz Frames.  The semi-classical model works best in the
string rest frame.  The quark and antiquarks materialize semiclassically,
simultaneously, at rest, equidistant from the breaking point.  Tunneling
is completely
symmetric between $q$ and $\bar{q}$.  This symmetry is lost in any other
Lorentz  frame.  If the materialization of $q$ and $\bar{q}$ are no longer
simultaneous, the $q$ and $\bar{q}$ will be moving, possibly at different
speeds.  Not only will the tunneling process be more complicated (e.g.
quarks have to absorb kinetic energy in addition to rest mass energy from
the field) but it might not even  make sense.  If the $q$ and $\bar{q}$
do not materialize simultaneously we will have unshielded, dangling color
fields. [If $q$ pops out first it will saturate the color field from
$\bar{Q}$, leaving $Q$'s color naked  while it waits for the $\bar{q}$ to
appear!]. For these
reasons, we feel  that the tunneling model is a reliable physical model
for string breaking primarily in the string rest frame.  This has
consequences for the fragmentation.

The tunneling model for string breaking when applied in the string rest
frame,  indicates that fragmentation must continue beyond the first stage.
Does the same argument force the fragmentation to proceed to  third or
higher stages?  As we shall see, the kinematics are significantly
different at the second stage, so that there is no necessity of higher
stage fragmentation.  In this sense there is something special about the
second stage.

According to our arguments the tunneling model for string breaking at the
second stage should be  applied in the rest frame of the fireball of mass
$m_1$.  We must therefore make a  Lorentz transformation from the lab frame
to the rest frame of $m_1$.  This Lorentz transformation is determined by
 
\be
\gamma = \frac{E_{m_{1}}}{m_{1}}
~~~\gamma\beta = \frac {P_{m_{1}}}{m_{1}}
\label{4.7}
\ee

The role of $P_0$ will now be played by Lorentz transformation of
$P_Q$,~$P^{1}_{Q}$
\be
P^{1}_{Q} = \Lambda P_Q
\label{4.8}
\ee
where $\Lambda$ is the Lorentz Transformation of (\ref{4.7}).
From (\ref{2.9})
\be
P_Q=P_0-\frac{\alpha}{\sqrt{2}} (x_++x_-)_{m_{1}}.
\label{4.9}
\ee
 For all cases of interest the $\bar{q}$, which has tunneled out of the
vacuum, will have asymptoted  to the light cone of a massless $\bar{q}$
produced at $x_1, t_1$ (See Fig.4). So $x_{m_{1}}^{-}= x_{1}^{-}$.
$x_{1}^{-}$ is
determined in terms of $m_{1}^{2}$ and $z_{1}$, (see (\ref{2.14}) 
\be
\frac {m_{1}^{2}}{z_{1}}-\mu^2
= (\sqrt{P_{0}^{2}+\mu^{2}}) \sqrt{2}\alpha x^-.
\label{4.10}
\ee
 $x_{m_{1}}^{+}$ is fixed by the equation of motion Eq.(\ref{2.5})
\be
x_{m_{1}}^{+} = 
\frac {x^{-}(P_{0}+\sqrt{P_{0}^{2}+\mu^{2}})}
{\frac{m_{1}^{2}}
{(P_0+\sqrt{P_{0}^{2}+\mu^{2})}
z_{1}}}
\simeq
\frac{\left (1-\frac{\mu^{2}z_{1}}{m^{2}_{1}}\right )}
{\sqrt{2}\alpha}
(2P_0)
\label{4.11}
\ee
where we assume $P_0\gg\mu$. Hence
\be
P_Q = P_0 z_1 \left ( \frac{\mu^2}{m_{1}^{2}}\right )
\label{4.12}
\ee
The kinematic region of greatest interest with $z_1$ relatively 
large corresponds to 
\be
\frac{z_{1} P_0\mu} {m_{1}^{2}} \geq 1
\label{4.13}
\ee
In this regime
\be
P^{1}_{Q} \simeq \frac{1}{2} \left (1+ \frac{\mu^2}{m_{1}^{2}}    \right)
m_1
\label{4.14}
\ee
which is much smaller than $P_0$ for all relevant values of $m_{1}$.
From Eq. (\ref{4.5}) we see that it is the large value of $P_0$ which
drives the fragmentation to large mass states, requiring further
fragmentation.  $P_Q$ (\ref{4.14}) will not become large unless $m_1$
becomes large.  But because of the exponentially rapid falloff with
$m_{1}^{2}$ of the first stage fragmentation function (\ref{2.17}) very
little fragmentation occurs with large $m_{1}^{2}$.

At the second stage the minimal mass $M^2$ (\ref{4.5}) becomes
\be
M^2=\mu^2+2m_\ell \sqrt{(P^{1}_{Q})^{2}z_{2}^{2}+\frac {\mu^{2}}{2}}
= \mu^{2}+2m_\ell 
\sqrt{\frac{\left (1+\frac{\mu^{2}}{m_{1}^{2}}\right
)}{4}^{2} m^{2}_{1} z_{2}^{2} + \frac{\mu^{2}}{2}}.
\label{4.15}
\ee

Again requiring that $M \geq 2.15~ GeV = 1.43 \mu$ we find that we need
$$
m_1 \geq (3-5) \mu
$$
to form a fireball at the second stage, sufficiently massive to force a
third stage of the fragmentation process. These massive, first stage
states, 
are almost never produced because of the $e^{-m^{2}_{1}/\mu^{2}}$ factor
in the first stage fragmentation function.

Our conclusion is that there is something special about the second stage
of the string fragmentation process.  The Bowler function which describes
first stage fragmentation is sharply peaked near $z=1$, and $m_1 \approx
\mu$.  When we account for the finite mass of the constituent quarks that
pop out of the vacuum to break the string  we find that only massive
states, with $z>1/2$, are abundantly produced.  These states must continue
to fragment at the second stage.  Except for extremely massive states,
fragmentation will end (i.e. produce either $D_0$ or $D_{0}^{*}$) at this
stage.  Third stage fragmentation can only be populated by first stage
fragments of such high masses that we expect very few of them to be 
produced.  The vast majority of $D$ and $D^*$ observed  will have been 
produced by two stages of breaking and will therefore be described by
the phenomenologically successful Eq. (\ref{3.20}). 

\sxn{{\bf Conclusions}}

The string model is widely invoked to provide physical insight into the
physics  of confinement in QCD.  Its relevance to hadrons production in
$e+, e-$ collisions 
has long been appreciated.  It has therefore been disappointing that a
characteristic feature of the experimental data, a suppression of heavy
meson production near $z \rightarrow 1$  
has defied simple  explanation in the string model.  Our main
accomplishment in this paper is the demonstration that the string model of
Artru, Mennessier and Bowler, when properly extended, gives rise in a very
natural way to a $(1-z)$ factor in heavy quark fragmentation.  In the
heavy
quark fragmentation studies he pioneered,  Bowler  somewhat
arbitrarily cut off the fragmentation process after a heavy  quark
mesonic system
formed.
We removed this restriction and studied in detail the subsequent
fragmentation  of this mesonic system.  The mesonic
system decays in a
cascade pattern, spitting out light mesons while degrading the fractional
momentum carried by the heavy quark meson.  Thus, at each stage of
fragmentation,  it is increasingly less likely that the heavy meson will
carry all the initial momentum.  It can not have $z=1$.  The calculation
makes this explicit, manifesting the degradation as a factor of $(1-z)$
for each stage beyond the first.  We found that the second stage
fragmentation function provided an excellent fit to experimental charm
fragmentation data and provides the rational for the heretofor 
mysterious
$(1-z)$ factor appended to the Bowler fragmentation function.
We then attempted to justify the predominance of the second stage
of the
fragmentation process.

An examination of the fundamental string  breaking mechanism, quark
anti-quarks tunneling  from the vacuum in a strong QCD Field, revealed the
non-Lorentz invariant nature of this process.  A preferred reference
frame,
the rest frame of the string, 
emerges as the natural stage on which to perform this quintessential
quantum act.   Redoing the fragmentation
analysis in this frame, which corresponds to the laboratory frame, the
existence of a non-zero effective quark mass for the {\em popped}  quarks
imposes a minimal mass for the heavy meson fragments produced. For all,
except the smallest $z$ values where fragmentation is unlikely in any
event, this minimal mass is well above the mass of {\em stable} heavy
quark mesons, implying that at least one more fragmentation stage is
necessary.  This is the physical reason why the first stage fragmentation
process is not relevant for the experimental observation.

This minimal mass effect is much less robust at higher stages of
fragmentation.
The Lorentz Transformation to the rest frame of the newly produced heavy
quark mesonic system greatly deflates the strength of this effect.  The
minimal mass system produced at the second stage is within the range of
stable heavy meson states.  Thus the second stage fragmentation function
is special.  Fragmentation must proceed to at least this stage, but can
end at this stage.  The phenomenological success of the modified Bowler
function can now be recognized as a success of the second stage of 
string
fragmentation.
\vglue 0.6cm
\noindent {\bf Acknowledgements}
\vglue 0.5cm

We wish to acknowledge partial support from DOE under contract number
DE-FG02-85-ER40231. We have benefitted from conversations with
encouragements from N. Horwitz and G.G. Moneti

\newpage
\noindent {\bf Figure Captions}
\vglue 0.6cm
\begin{description}

\item[Fig. 2] First stage heavy quark fragmentation in the laboratory
frame.  The string joining $Q_0$ to $\bar {Q}_0$ breaks at $(x_1,t_1)$ by
producing a massless $q \bar{q}$ pair. The $\bar{q}$ is then accelerated
by the string attaching it to $Q_0$.  The $\bar{q}$ and $Q_0$ trajectories
intersect at point $m_1$ terminating first stage fragmentation with the
production of a fireball of mass $m_1$.

\item[Fig. 2] Second stage fragmentation in the heavy quark rest frame.
The first stage terminated at point $m_1$.  Second stage fragmentation is
caused by the string breaking at $(x_2,t_2)$ and terminates when $\bar
{q}_2$ intersects the world line of $Q_0$ at $m_3$, producing a heavy
quark mesonic state of mass $m_3$.  The shaded region in the area we call
$A_2$ in section 2.

\item[Fig. 3] A comparison between the modified Bowler function of Ref. 6
to our second stage fragmentation function.  By a slight shift in a
phenomenologically determined parameter (B= .65 rather than .63) we see
that the
two functions are essentially indistinguishable.

\item[Fig. 4] String breaking by the production of a massive $q,\bar{q}$
pair.  The $q$ and $\bar{q}$ no longer appear at the same location.  As
the $q$ and $\bar{q}$ accelerate their worldline eventually asymptote to
the
light like trajectories of a massless $q$ (or $\bar{q}$)y
 produced at $x_1
t_1$.

\end{description} 

\end{document}